\documentclass[pra,twocolumn,showpacs,preprintnumbers,amsmath,linenumbers,amssymb,superscriptaddress]{revtex4}
\usepackage{amsfonts}
\usepackage{mathrsfs}
\usepackage{amsmath}
\usepackage{amssymb}
\usepackage{txfonts}

\usepackage{graphicx}
\usepackage{dcolumn}
\usepackage{bm}
\usepackage[dvipdfm]{xcolor}
\usepackage{subfigure}

\begin{document}



\title{Roughness with a finite correlation length in the Microtrap }

\author{Muzhi Wu}
\affiliation{School of Electronics Engineering $\&$ Computer
Science, Peking University, Beijing  100871, China}
\author{Xiaoji Zhou}\thanks{Electronic address: xjzhou@pku.edu.cn}
\affiliation{School of Electronics Engineering $\&$ Computer
Science, Peking University, Beijing 100871, China}
\author{W. M. Liu}
\affiliation{Institute of Physics, Chinese Academy of Sciences,
Beijing 100080, China}
\author{Xuzong Chen}
\affiliation{School of Electronics Engineering $\&$ Computer
Science, Peking University, Beijing  100871, China}


\begin{abstract}
We analyze the effects of roughness in the magnitude
of the magnetic field produced by a current carrying microwire,
which is caused by geometric fluctuation of the edge of wire.
The relation between the fluctuation of the trapping
potential and the height that atom trap lies above the wire is
consistent with the experimental data very well, when the colored
noise with a finite correlation length is considered.
On this basis, we generate the
random potential and get the density distribution of the BEC atoms
by solving the Gross-Pitaevskii equation, which coincides well with
the experimental image, especially in the number of fragmentations.
The results help us
further understand the nature of the fluctuation and predict the
possible application in the precise measurement.
\end{abstract}


\maketitle

\section{INTRODUCTION} Trapping and manipulating ultracold atoms
and Bose-Einstein condensations (BEC) in magnetic potential
produced by micro atom chip attracted more attention recently because it is easy to perform
coherent manipulation, transport and interferometry on a
chip~\cite{Folman,Fort2007}, where the condensate is  very close to the surface of wire.
It is possible to bring miniaturization into the application of BEC~\cite{Hansel}.
The trapping frequencies in the radial direction are normally several $k Hz$, and a few $Hz$ in the axial direction. Hence the
aspect ratio is so high that the size of BEC is several $mm$ in the axial direction, the spatial effects on the
condensate will be obvious. However, for small-scale magnetic field of atom
traps produced by micro fabricated current-carrying wires, an
unexpected problem in these chips was their small fluctuation in
magnetic field, which introduces a random potential along
the trap. Such a fluctuation in potential will cause
fragmentation of the atom clouds~\cite{Fort2007,Leanhardt}. The
effects of the random potential have been studied
theoretically~\cite{Wang,Schumm}, where the randomness in magnetic
field is due to geometric fluctuation of the wire surface and the
strength of the disorder field shows a $d^{-5/2}$ relation with the
height $d$ that atoms lie above the wire, when assuming the correlation
of the wire edge is the white-noise correlation. It is not
coincidental to the experimentally observed $d^{-2.2}$
relation~\cite{Kraft}.

The deviation in the theoretical calculations~\cite{Wang,Schumm} from
the experiment is caused by the white-noise approximation made,
which does not meet the actual situation that the system's fluctuation
cannot be the ideal white noise~\cite{palasantzas, hinds}. In this
paper we elucidate the nature of the random potential, indicating
the relation between the strength of the potential and the height of
the BEC atoms. Different from the white-noise assumption, we consider a more general correlation form---the colored noise
correlation $R(r)=\sigma^2e^{-|r|/\xi}$,
with $\xi$ the correlation length of the system.
Here we borrow the notion 'colored noise' from the theory for time-domain noise
analysis to deal with the case of spatial roughness~\cite{Zhou}.
Our theory shows that the power index $\alpha$ in the disorder potential $u_s=C\cdot d^{-\alpha}$
is relevant
to the system's intrinsic correlation length $\xi$. This means by
measuring the dependence factor $\alpha$, the correlation length
of the current-carrying wire can be known.  We point out in experiment
~\cite{Kraft}, $\alpha=2.2$ because $\xi=75\mu m$.
Consequently, the correlation function for the random
potential is calculated.

\begin{figure}
\label{fig:setup}
\includegraphics[width=7cm]{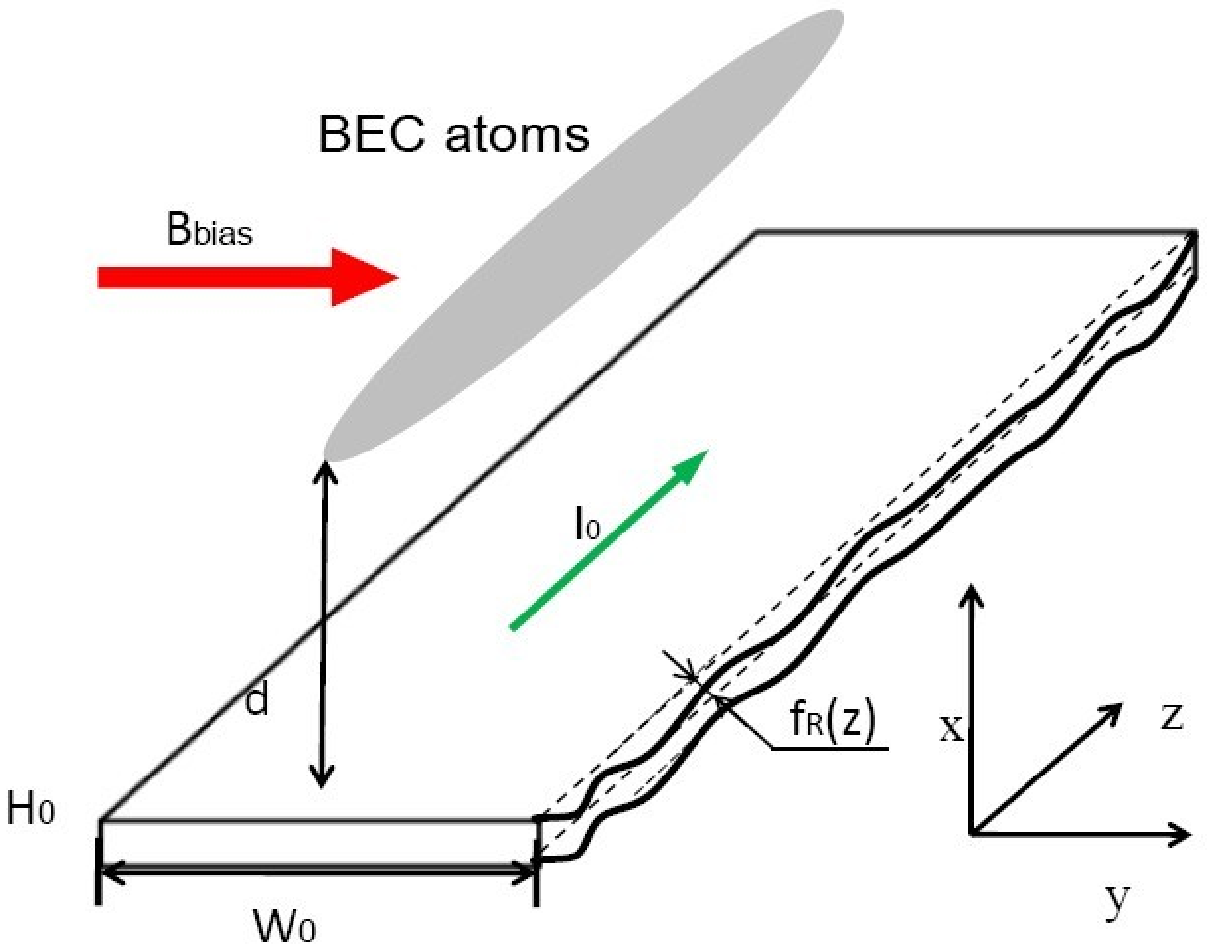}
\caption{(Color online) A typical atomchip for BEC.  A constant current
$I_0$ flows in the z-direction of the wire with $H_0$ high and $W_0$ wide. The BEC atoms
lie above the wire at the height of $d$. A bias magnetic field $B_bias$ in
the longitudinal direction produces a local minimum. $f_R(z)$ is
the deviation of the edge from its ideal position.}
\end{figure}

With the random potential we solve the Gross-Pitaevskii equation, depict the
density profile of the captured BEC atoms and compare it with that of the white-noise case.
The desired random potential is generated using a selected
aperture function derived from the correlation function.
The number of fragmentations has a better agreement with the
experiment than previous calculations. This work is meaningful in
the simulation of transport process. It as well predicts a criterion
for judging the quality of the wire used for cold-atom manipulation,
which might be highly applicable in precise measurement.

\section{THE STRENGHTH OF THE DISORDER POTENTIAL}
A typical setup of a magnetic microtrap is shown in Fig.~1. The current $I_0$ flowing in a micro-fabricated copper conductor
in the z-direction and a bias field $B_{bias}$ in the y-direction form a magnetic trap for
atoms. The distance $d$ from the surface of the current-carrying wire to the center of the
trapping potential is controlled by the bias field and the current,
$d=\mu_0I_0/2\pi B_{bias}$.

The random potential is brought about by the roughness in the
magnetic trap for the BEC atoms lying above the wire surface about
$100\mu m$ $\sim$ $200\mu m$. What we are interested
 is the z-direction of the random magnetic field $B_z$, which connects with
 the y-component of the current. In experiments~\cite{Esteve,Kraft}, the wire normally is a thin
electroplated gold layer of a few $\mu m$ thick.  The surface of the wire can
be made very flat technically with the electroplating method. So the
main contribution of the random field arises from the side of the
wire.

The distortion of the side edge of the wire curves the current flow.
The potential roughness in the magnetic microtrap is induced by
this abnormal current. To compute the fluctuation without loss of
generality, we make the following assumptions. First the current
$I_0$ remains constant on the edge of the wire, the amplitude $J_0$ of the current
density $\vec j$ is then constant too. Second the
resistivity keeps unchanged along the current-carrying wire. Third
the deviation of the edge from its ideal position (both left and
right) $f_{L/R}$ is trivial compared with the width of the wire
$W_0$, $f_{L/R}/\ll W_0$.
Considering symmetry, only $f(z)\equiv\frac{1}{2}(f_L(z)+f_{R}(z))$ contributes notably to the fluctuation,
and $f^-=\frac{1}{2}(f_L-f_R)$ is neglected.

If we ignore the change in the module of the current density vector, then we consider the effects
due to the alteration of the direction of $\vec j$. The
z-component of $\vec B$, i.e. $B_z$, is directly relevant to
the y-component of current density.
In yz plane the current density satisfies the charge conservation condition
$\frac{\partial j_y}{\partial y}+\frac{\partial j_z}{\partial z}=0$.
Under such assumptions, we introduce an auxiliary scalar potential
$G$, so that $j_y=J_0\frac{\partial G}{\partial z}$ and
$j_z=-J_0\frac{\partial G}{\partial y}$. The function $G$
satisfies the Laplace equation in the interior of the wire $\nabla^2
G=0$. Using separation of variables, we get the Fourier
component of $f(z)$ which brings a transverse current
density~\cite{Wang},

\begin{equation}
j_y\propto J_0\frac{\partial}{\partial z} \int_{-\infty}^{\infty}
\frac{e^{ikz}dk}{\sinh(kW_0)}\cdot\cosh(ky)\sinh(kW_0/2)F_f
\end{equation}
where $F_f$ is the inverse Fourier transform of $f(z)$,
$F_f=\frac{1}{2}\int_{-\infty}^\infty dze^{ikz}f(z)$, and $k$ is
the wave vector in the frequency domain.

Therefore the disorder magnetic field can be
derived from the Bio-Savart law:
\begin{equation}
\vec B(\vec r)=\frac{\mu_0}{4\pi}\int d^3 \vec r\frac{\vec{j}\times
\vec{r}}{r^3}
\end{equation}

The random potential $U(z)$ is given by $U(z)=-\mu_B\cdot B_z$. So
the strength of $U$ induced by the roughness in the magnetic
microtrap can be measured by the value at the zero point of the
potential's auto-correlation function $g(r)$ with $r=z-z'$,
\begin{equation}
g(r)=\langle \delta U(z)\delta U(z') \rangle
\end{equation}
$\langle\ldots\rangle$ denotes the ensemble average.
And the Fourier transform of $g(r)$, i.e. the power spectral density
of the random potential, is
\begin{equation}
S(k)=\int dr e^{ikr}g(r)
\end{equation}
As the wire is eletroplated with a thin layer of gold~\cite{Esteve},
it is acceptable to assume $H_0\ll d$ and ignore the thickness of
the line. We substitute the dimensionless variable $\beta=kd$ for $k$
, and obtain
\begin{equation}
S(\beta)=A d^{-4}\cdot \beta^4 P(\beta)
\end{equation}
with
\begin{eqnarray} \nonumber P(\beta)=\frac{2\sinh(\beta
W_0/2d)}{\sinh(\beta W_0/d)(\beta
W_0/d)}\sum_{n=0}^\infty\frac{(-1)^n}{n!(2y)^n}K_{n+1}(y)\cdot\\\gamma(2n+1,-\frac{\beta
W_0}{2d},
        \frac{\beta W_0}{2d})\cdot F(\beta/d)
\end{eqnarray}
where $A$ is a constant, $F(k)=\int d(z-z')e^{-ik(z-z')}\langle
f(z)f(z')\rangle$, $K_n(x)$ the Bessel function of the
second kind, and $\gamma(n,x_1,x_2)$ is the incomplete gamma
function $\gamma(n,x_1,x_2)=\int_{x_1}^{x_2}dx x^{n-1}e^{-x}$. We
then compute the inverse Fourier transform and get the
disorder potential $u_s^2$.
\begin{equation}
g(z)=\mathcal{F} ^{-1}[S(\beta)]=\frac{1}{d^5}\int_{-\infty}^\infty d\beta e^{ik(z-z')}S(\beta)
\end{equation}
\begin{equation}
\label{eq:us}
u_S^2=g(0)\sim d^{-5}\int_{-\infty}^\infty dkF(k)
\end{equation}

When the correlation of the wire edge fluctuation
is white-noise form, $\langle f(z)f(z')\rangle$ is a  $\delta$ function,
and $F(k)$ in Eq.(8) is a constant, we have $u_s^2\sim d^{-5/2}$, which is
the same in~\cite{Wang}. The white noise
condition when the edge of the wire totally not correlated is the
simplified case, it limits the supremum of the system fluctuation. Below we discuss the colored-noise
model with a finite correlation length.

\section{DEPENDENCE BETWEEN POTENTIAL AND HEIGHT FOR A FINITE CORRELATION LENGTH}

The fluctuation $f(z)$ is described as a colored noise model. The function $f(z)$ is a Gaussian random
variable with zero mean $\langle f(z)\rangle=0$, and variance~\cite{palasantzas, hinds}
\begin{equation}\label{eq:colored}
\langle f(z)f(z')\rangle = \sigma^2e^{-|r|/\xi}
\end{equation}
The colored noise given by Eq.(\ref{eq:colored}) is parameterized by three key factors,
the distance along the wire $r=z-z'$, the characteristic correlation length $\xi$  and the noise strength $\sigma$.
When $\xi\rightarrow 0$, it degenerates into the white-noise correlation form $\langle f(z)f(z')=\delta(z-z')$,
which is the ideal situation of the roughness~\cite{Wang,Schumm}. When $\xi\rightarrow\infty$, the surface of the wire becomes totally correlated,
implying a perfect smooth surface, which is the ideal case of the wire. The power spectrum of Eq.(\ref{eq:colored}) for the  wire edge is
\begin{equation}
\label{eq:fk}
F(k)=\frac{2\xi}{1+\xi^{2}k^{2}}
\end{equation}

\begin{figure}[!h]
\centering
\label{fig:dk_t}
\includegraphics[width=6cm]{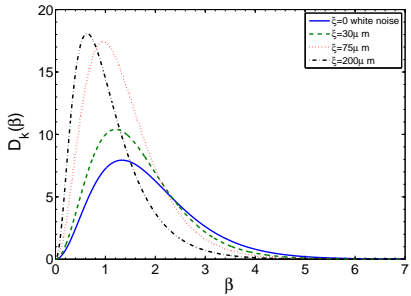}
\caption{(Color online) The power spectral function of potential $S(\beta)$ for the different correlation length of disorder potential.
The solid line is for white noise with $\xi=0 \mu m$. Others are for the case of colored noise.
The dashed line with $\xi=30\mu m$, the dotted line $\xi=75\mu m$, and the dash dotted line $\xi=200\mu m$. }
\end{figure}

When it comes to our model that the wire edge has an
intrinsic randomness, the power spectral density of which
described in Eq.(\ref{eq:fk}) is moderate in comparison.
The power spectral density of potential
$S(\beta)$ for different correlation lengths is shown in Fig.~2. In the white-noise
correlation $\xi=0 \mu m$, the curve peaks at $\beta=1.3$.
If we choose the typical correlation length $\xi=75\mu m$~\cite{Kraft},
the curve peaks at $\beta=1.1$. From
Fig.~2, we know the power spectral density peaks at a smaller value $\beta$ with the
increasing $\xi$. This means wave vector $k$ decreases and the periodicity weakens for
fluctuation at a certain $d$. The height of the peak goes up as the correlation
length increases, which shows qualitatively the wire edge becomes
smooth so that the frequency region is smaller for disorder potential.

In order to clarify the effect of the wire width $W_0$,
we plot the relation between the random potential $u_s$ and the
ratio $d/W_0$ in logarithmic coordinates in Fig.~3. To express it clearly, we can
divide the figure into three regions. For distance $d$ varying from 0 to $W_0$, the BEC atoms
are very close to the top surface of the wire. The relation between the random potential and the
ratio is a linear dependence, which is the fat-line approximation. When $W_0<d<3W_0$,
the situation becomes complicated, the rim effect
of the wire as well as the orbiting magnetic field combine to
present a field that cannot be easily calculated theoretically. For distance $d$ ranging from $3W_0$ to
infinity, this is also a linear fit corresponding to the thin-line condition. The BEC atoms are far from
the top, so the width of the wire can be neglected and the wire as a whole dominates the field.
In common experimental conditions~\cite{Kraft,Esteve}, the atoms are placed high
above the surface of the wire with $W_0$ neglected.
So below, during our discussion on the $\alpha\sim\xi$ relation, the parameter $d$
we choose mainly ranges from $100\mu m$ to $200\mu m$.

\begin{figure}[!htb]
\label{fig:usd}
\begin{center}
\includegraphics[width=6cm]{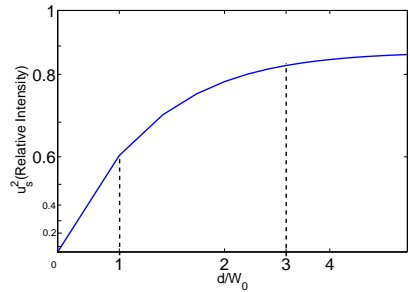}
\end{center}
\caption{(Color online) The scaled power spectral density of the random potential and the
dimensionless factor $d/W_0$ in logarithmic coordinates. In the area
of $0<d/W_0<1$ and $d/W_0>3$, the curve has a good linear fit.}
\end{figure}

The dependence between the strength of the random potential and $d$ can be assumed
as
\begin{equation}
\label{eq:alpha}
u_s=C(d)\cdot d^{-\alpha}
\end{equation}
Based on the above analysis, the distance that the atoms lie above the wire $d$
have a linear fit in the logarithmic coordinates for the thin-line approximation when $d\gg W_0$ ,
so $C(d)$ is a constant describing the amplitude of the random fields,
$\alpha$ the power index characterizing the speed the fields decays
with $d$. The strength of the disorder potential drops with the
growth of height $d$.

Using Eq.(\ref{eq:us}) and Eq.(\ref{eq:alpha}), we calculate $u_s$
for different $\xi$, and fit them against $d$ to get $\alpha$.
The relationship between $\alpha$ and $\xi$ is plotted in Fig.~4.
We can see the potential roughness decays with the height $d$ in direct relation to $\alpha$.
As the correlation length $\xi$ increases, the index $\alpha$ decreases drastically.
When $\xi$ is zero, $\alpha\rightarrow 2.5$, which is coincidental to the
white-noise approximation.
Provided that the deviation between the
data from the experiment and theoretical calculation is mainly
caused by the uncertainty of $\xi$, then in our calculations we
elucidate the correlation length of the system, i.e., the
fluctuation of the edge of the wire should be a finite number on a
large scale. For instance, the correlation length in
experiments~\cite{Kraft,Esteve} should be $75\mu m$ pointed out by
our figure as well as further calculations.
The power index $\alpha$ is determined as soon as $\xi$ is
determined.

The strength factor $\sigma^2$ in the colored-noise correlation function Eq.(9) only contributes to the constant before the integral in
Eq.(7) and is totally not relevant to the shape of the random potential,
so it has no effect on the power index $\alpha$. However we can see below its influence on the density
profile is similar to the alteration of the distance $d$.

\begin{figure}[!h]
\begin{center}
\includegraphics[width=6cm]{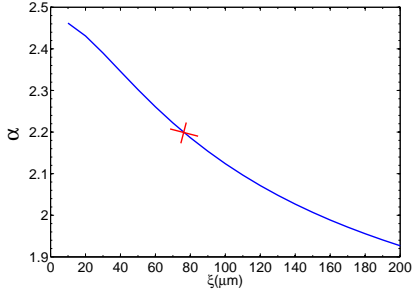}
\end{center}
\caption{(Color online) Relationship between the power index $\alpha$ and the
correlation length $\xi$ in the case of thin-line approximation ($d\gg W_0$). The
cross point is the typical value $\xi=75\mu m$ and $\alpha=2.2$.}
\end{figure}

\section{THE FORM OF ROUGH POTENTIAL AND FLUCUATIONS OF THE DISTRIBUTION}
We then discuss the disturbance of the distribution of BEC atoms caused by the random magnetic fields.
The rough potential in the trap can be obtained in experiments by measuring the density profile of cold
atom clouds via absorption imaging method.
As the BEC atoms expand mainly in the z-direction, and are very tight in the other two dimensions,
the distribution can be described by the one-dimension Gross-Pitaevskii Equation(GPE)~\cite{Dalfovo}
\begin{equation}
i\hbar\frac{\partial\Psi}{\partial
t}=-\frac{\hbar^2}{2m}\nabla^2\Psi+V(z)\Psi
+NU_0\left|\Psi\right|^2\Psi
\end{equation}
$\Psi$ satisfies the unitary normalization condition, $N$ is the number of BEC
atoms and $U_0=4\pi\hbar^2 a/m$ is the intensity of mean field interaction.
$V(z)$ includes both harmonic potential and the fluctuated potential as
follows~\cite{Bao}
\begin{equation}
  V(z)=\frac{1}{2}m\omega_z^2 z^2+V_d(z).
\end{equation}
$V_d(z)$ is the random
potential induced by the fluctuation, and $\omega_z$ is the frequency in the z-direction.
Here we use a method similar
to~\cite{Modugno} to generate a random potential with desired
correlation function.
\begin{equation}
V_d(z)=\mathcal F^{-1}[W(k)\mathcal F[g(z)]],
\end{equation}
where $W(k)$ is a an appropriate aperture function which can filter out high and low part of the signal. With $V_d(z)$ we use
backward euler finite difference (BEFD) to
compute the dynamic evolution of the GPE and finally get the wave function $\Psi$ and the density profile $|\Psi|^2$~\cite{Bao}.

Due to the random potential induced by the fluctuation of the
magnetic fields, a series of fragmentations in the condensate
density profiles come into being. For different bias field values,
the experimental sequence used to produce ultra-cold atoms in this
trap were carried out, corresponding to different distances from the
magnetic trap to the top surface of the wire. The strength
of the random potential is mainly affected by the value
$\sigma^2$ in Eq.(9). When the strength is relatively large, the
random potential is comparable to the harmonic potential.
The fragments present a high peak, and the density distribution
rises and falls sharply, as shown in Fig.~5.

\begin{figure}[!h] \label{fig:xi}
\includegraphics[width=8cm]{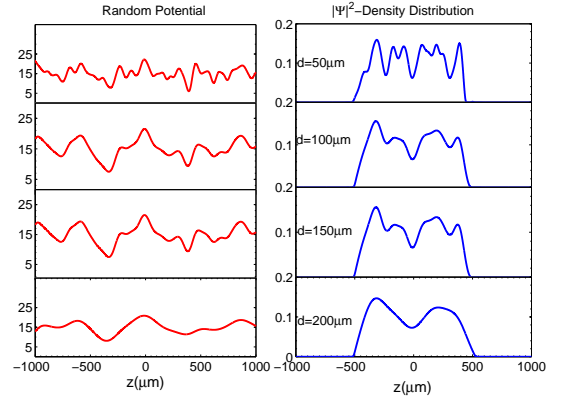}
\caption{(Color online) Density profile of the BEC atoms and the random potential
at different heights on the case of large random potential.
The left column indicates the random potential
at 50$\mu m$, 100$\mu m$, 150$\mu m$ and 200$\mu m$ respectively from top to bottom.
The right column is for the
normalized distribution $|\Psi|^2$ at the height of 50$\mu m$, 100$\mu m$, 150$\mu m$ and
200$\mu m$ from top to bottom. Figures for $|\Psi|^2$ range the same from 0 to 0.2.
$\xi$ is fixed at $75\mu m$.~\cite{Kraft}}
\end{figure}

Fig.~6 shows effect of the disorder strength on the density profile.
We focus on the area around the main peak and take out the $1000\mu m$ area.
The relative roughness strength in Fig.~6~(a-d) is 1:2:6:10.
From the figure we can see the fragmentation gradually becomes obvious
as the strength increases. At the appropriate strength
for the random potential, as shown in Fig.~6(c), the density profile
basically coincides with the experiment~\cite{Kraft}.
The amplitude for the fragments are different with
one highest and others decreasing accordingly, and
the distance between two peaks in the $1000\mu m$ area
is about $250\mu m$. The characteristics of the experiment~\cite{Kraft}
are reflected by our calculations.

\begin{figure}[!h] \label{fig:weak}
\includegraphics[width=8cm]{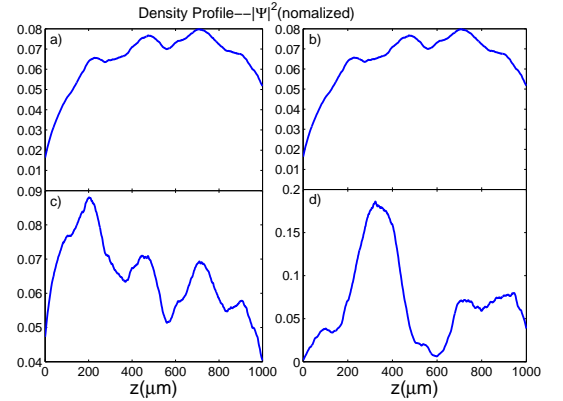}
\caption{(Color online) Effect of the disorder strength on the density profile
with $\xi=75\mu m$. The relative strength in (a), (b), (c) and (d) is 1:2:6:10.}
\end{figure}

First we discuss the effects of height $d$ at a finite correlation length $\xi$.
We can see the dependence between the random potential and the distance in the left column of Fig.~5. It
indicates the random potential at 50$\mu m$, 100$\mu m$, 150$\mu m$ and 200$\mu m$.
The right column is for the density distribution.
Using the equation $u_s^2=Cd^{-2.2}$ given above, we
can assess the strength of the potential quantitatively.
As the distance increases, the roughness of the
random potential weakens.
Meanwhile the fragmentation phenomenon in the density
distribution becomes less obvious.

\begin{figure}[!h] \label{fig:d}
\includegraphics[width=8cm]{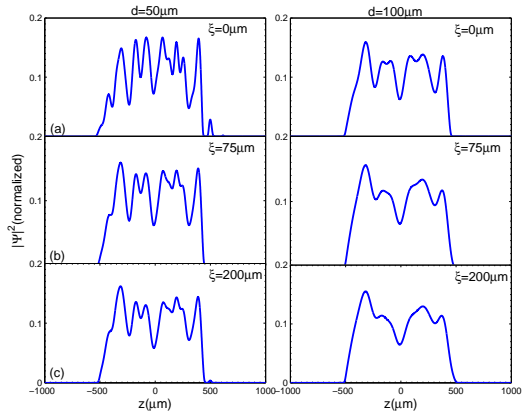}
\caption{(Color online) The normalized density profile of the BEC atoms for different correlation
length. The left column for the height $d=50\mu m$. The right column
for $d=100\mu m$. All the figures
have the same range for $|\Psi|^2$ from 0 to 0.2.
 (a) $\xi=0\mu m$ (b) $\xi=75\mu m$ (c) $\xi=200\mu m$. }
\end{figure}

Then we show the influence of different correlation length at a fixed height $d$.
Fig.~7 demonstrates how the correlation length of the
system affects the distribution of BEC atoms. The left column is for
different correlation lengths 0$\mu m$, 75$\mu m$, 200$\mu m$ at the height $d=50\mu m$, and the right
column is for all the same conditions except for $d=100\mu m$.
When the system has an intrinsic correlation length, the fragmentation of the
density profile is apparently small than that of the white noise. In
two extreme cases, as $\xi\rightarrow\infty$, the distribution
becomes smooth because it is the ideal situation that the
current-carrying wire has no fluctuation, while
$\xi\rightarrow 0$ is the simple white-noise case, and the
fluctuation is at its maximum.
We are most interested in Fig.~6(c) when $d=100\mu m$ and
$\xi=75\mu m$. This set of data is measured in the experiment~\cite{Kraft}.
The number of the fragmentations, which is about 4 along the z-axis over
$1000\mu m$, is in good agreement with the image derived
from the experiment, better than that of the white-noise approximation (about 6)
which is different from the experimental data.

\section{DISCUSSION AND CONCLUSION}

According to our analysis, different correlation length will lead to
different $\alpha$. As in experiment~\cite{Kraft,Esteve},
$\alpha=2.2$, the plausible correlation length through our
calculations, should be $75\mu m$, and the density profile we thus
get is coincidental with the experiment. The spectral density of the
distribution~\cite{Esteve} shows that $\xi$ has a very small
fluctuation $100nm$, so the features of a system could be considered
as a  finite correlation length. We also take the width of the wire
into account, pointing out exactly how it affects the random
potential. Though the real trap is not a thin-wire, our model
indicates when the height $d$ is relatively large, it is rational to
ignore the width of the wire. We then analyze how the strength of
the random potential affects the density distribution. The density
profile for the appropriate disorder strenth in our theoretical
picture is in better accordance than the previous white-noise
approximation to some extent, which proves our conclusions to be
more reasonable. That is to say, by counting the number of
fragmentations of the atoms, we can see qualitatively the
correlation length of the wire. By furthermore measuring the power
spectral density of the disorder potential, we can get the
correlation length quantitatively.

In summary, we bring forth a more general model to calculate the
random potential for the BEC atoms in a current-carrying wire. We
show that the fluctuation in the potential is caused by the
distortion of the wire edge with a fixed correlation length.
The correlation length of the specific wire used in experiment~\cite{Kraft}
is $\xi=75\mu m$ and the power index in $u_s=C\cdot d^{-\alpha}$ is
$2.2$. This might be applicable in the precise measurement because of the
possible criterion provided for the necessary wire quality to be
used for cold atom manipulation. Besides, by measuring the spectral density
of cold atom distribution and analyzing the deviation in the data we
get from theoretical calculations and experiment, we can get to know the
nature of the noise, and apply it in the simulation of the transport process.

\section{ACKNOWLEGEMENT}
X. J. Zhou thanks to E. A. Hinds, B. Darquie and H. T. Yin for their help and discussion.
This work is partially supported by the
state Key Development Program for Basic Research of China
(No.2005CB724503, 2006CB921401, 2006CB921402), and by NSFC (No.10874008 and 10934010).

\end{document}